\documentclass[pss]{wiley2sp} 
\usepackage{amsmath}

\newcommand{\unit}[1]{\ensuremath{\, \mathrm{#1}}}
\newcommand*{\fg}[1]{Fig.\thinspace\ref{#1}}
\newcommand*{\fgs}[1]{Figs.\thinspace\ref{#1}}
\newcommand*{\eq}[1]{Eq.\thinspace\ref{#1}}

\begin{document}

\title{Unambiguous determination of spin dephasing times in ZnO by time-resolved magneto-optical pump-probe experiments}

\titlerunning{Unambiguous determination of spin dephasing times in ZnO}

\author{%
  Sebastian Kuhlen\textsuperscript{\textsf{\bfseries 1},\textsf{\bfseries 2}},
  Ralph Ledesch\textsuperscript{\textsf{\bfseries 1},\textsf{\bfseries 2}},
  Robin de Winter\textsuperscript{\textsf{\bfseries 1},\textsf{\bfseries 2}},
  Matthias Althammer\textsuperscript{\textsf{\bfseries 3}},
  Sebastian T. B. G\"{o}nnenwein\textsuperscript{\textsf{\bfseries 3}},
  Matthias Opel\textsuperscript{\textsf{\bfseries 3}},
  Rudolf Gross\textsuperscript{\textsf{\bfseries 3}},
  Thomas A. Wassner\textsuperscript{\textsf{\bfseries 4}},
  Martin S. Brandt\textsuperscript{\textsf{\bfseries 4}}, and
  Bernd Beschoten\textsuperscript{\Ast,\textsf{\bfseries 1},\textsf{\bfseries 2}}}

\authorrunning{Kuhlen et al.}

\mail{e-mail
  \textsf{bernd.beschoten@physik.rwth-aachen.de}}

\institute{%
  \textsuperscript{1}\, II. Physikalisches Institut, RWTH Aachen University, 52056 Aachen, Germany\\
  \textsuperscript{2}\, JARA-Fundamentals of Future Information Technology, J\"{u}lich-Aachen Research Alliance, Germany\\
  \textsuperscript{3}\, Walther-Mei{\ss}ner-Institut, Bayerische Akademie der Wissenschaften, 85748 Garching, Germany\\
  \textsuperscript{4}\, Walter Schottky Institut and Physics Department, Technische Universit\"{a}t M\"{u}nchen, 85748 Garching, Germany}

\received{XXXX, revised XXXX, accepted XXXX} 
\published{XXXX} 


\abstract{%
\abstcol{%
Time-resolved magneto-optics is a well-established optical pump probe technique to generate and to probe spin coherence in semiconductors. By this method, spin dephasing times $T_2^*$ can easily be determined if their values are comparable to the available pump-probe-delays. If $T_2^*$ exceeds the laser repetition time, however, resonant spin amplification (RSA) can equally be used to extract $T_2^*$. We demonstrate that in ZnO these techniques have several tripping hazards resulting in deceptive values for $T_2^*$ and show how to avoid them.
  }{
 We show that the temperature dependence of the amplitude ratio of two separate spin species can easily be misinterpreted as a strongly temperature dependent $T_2^*$ of a single spin ensemble, while the two spin species have $T_2^*$ values which are nearly independent of temperature. Additionally, consecutive pump pulses can significantly diminish the spin polarization, which remains from previous pump pulses. While this barely affects $T_2^*$ values extracted from delay line scans, it results in seemingly shorter $T_2^*$ values in RSA.

 }}


\maketitle   

\section{Introduction}

For almost two decades ultrafast magneto-optical pump-probe methods have become standard techniques for triggering and probing spin coherence in direct band gap semiconductors \cite{PhysRevLett.72.717,PhysRevB.56.7574,PhysRevLett.77.2814,Kikkawa29081997,PhysRevLett.80.4313,Nature397_Kikkawa1999,PRB63_Beschoten2001_SpinCoherenceandDephasinginGaN,PhysRevB.63.235201,ghosh:232507,PRL96_Greilich2006,Greilich21072006,NaturePhysics3_Meier2007,PRB75_Schreiber2007,PhysRevB.82.155325,PhysRevLett.105.246603,PhysRevLett.107.137402}. Electron and hole spins are excited either by circularly \cite{PhysRevLett.80.4313} or linearly \cite{PhysRevLett.105.246603} polarized optical pump pulses. Spin precession in a transverse magnetic field is then usually monitored by linearly polarized probe pulses, either measuring the polarization rotation in transmission (Faraday effect) \cite{PhysRevB.56.7574} or reflection (Kerr effect) \cite{Kikkawa29081997}. Alternatively, their respective ellipticities can be analyzed \cite{PhysRevLett.107.137402}. These versatile techniques have been applied to a multitude of materials \cite{PhysRevLett.80.4313,PRB63_Beschoten2001_SpinCoherenceandDephasinginGaN,ghosh:232507,PSSB:PSSB200564604,PhysRevLett.95.017204,PhysRevB.76.205310} in bulk \cite{PhysRevLett.80.4313} and low dimensional geometries \cite{PhysRevB.56.7574,Kikkawa29081997,PRB75_Schreiber2007,PhysRevB.82.155325,PhysRevLett.107.137402}. It also allows for spatially-resolved spin detection \cite{Nature397_Kikkawa1999,Nature427_Kato2004,PRL94_Crooker2005} and can even be combined with time-resolved electrical methods to either probe spin precession after time-resolved electrical spin injection \cite{Schreiber} as well as after time-resolved spin polarization by electric field pulses \cite{PhysRevLett.93.176601} or to monitor spin rotations by either static or pulsed electric fields \cite{NaturePhysics3_Meier2007,Nature427_Kato2004,PhysRevLett.109.146603,PRB82_Norman2010}.

A key task in these experiments is the determination of spin dephasing times $T_2^*$ which are usually extracted from the exponential decay of the time-domain magneto-optical signal during spin precession. Their values can be precisely determined from pump-probe measurements with variable delay time \cite{PhysRevLett.80.4313} if they are in the order of the available pump-probe-delay. Mechanical delay lines typically cover up to 3~ns.

If $T_2^*$ exceeds the laser repetition period $T_{\text{rep}}$, resonant spin amplification (RSA) can equally be used to determine $T_2^*$ \cite{PhysRevLett.80.4313}. In this method, the external magnetic field is swept at a fixed pump probe delay. RSA manifests itself as sharp resonances in the magneto-optical signal whenever the laser repetition rate is in resonance with the spin precession frequency. The width of these resonances as a function of the external magnetic field is a direct measure of the respective spin dephasing times. Investigating spin dephasing in ZnO with unintentional aluminum and indium doping, we demonstrate that the above techniques have several tripping hazards resulting in deceptive results for $T_2^*$. We show that these pitfalls can be avoided to obtain unambiguous values for the spin dephasing times.

ZnO is particularly interesting for spintronics as it has a large bandgap and small spin-orbit coupling \cite{fu:093712}, promising long spin dephasing times. For this system both electrical spin-injection \cite{Chen200291,JI2009153,5467466,althammer:082404} and magneto-optical pump-probe experiments \cite{ghosh:232507,ghosh:162109,doi:10.1021/nl801057q,eigenes,eigenes2} have been studied, the latter demonstrating spin coherence up to room temperature. Here, we show that in ZnO the temperature dependence of the amplitude ratio of two separate spin species can easily be misinterpreted as a strongly temperature dependent $T_2^*$ of a single ensemble, while the two spin species have $T_2^*$ which only weakly depend on temperature. Furthermore, the fallback method RSA fails as well. Consecutive pump pulses can significantly diminish the spin polarization, which remains from previous pump pulses. While this barely affects pump-probe experiments in the time domain, it results in seemingly shorter spin dephasing times in RSA experiments. We demonstrate that an unambiguous determination of $T_2^*$ in ZnO require a careful analysis of the spin precession signal over the full laser repetition period.

\section{Experimental setup}

\begin{figure}[tb]%
\includegraphics*[width=\linewidth]{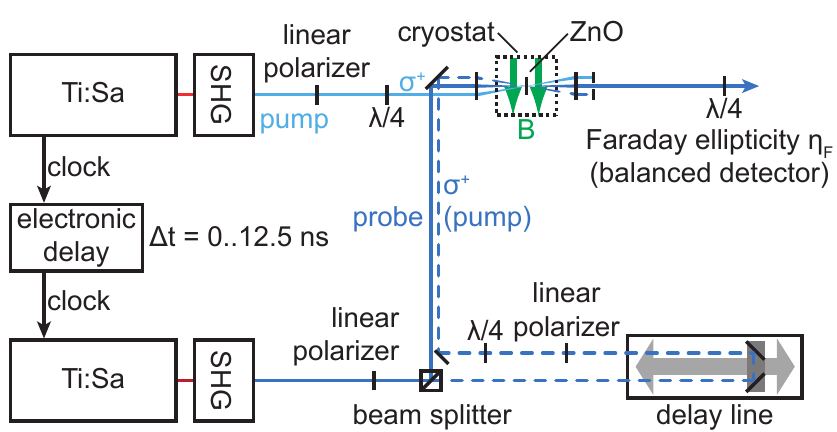}
\caption{Experimental setup. Pump and probe laser pulses stem from two independently tunable Ti-Sapphire ps lasers which are synchronized and electronically delayed. The dashed line represents a second pump pulse train, which is used in pump-pump-probe experiments.}
\label{fig1}
\end{figure}

The experimental setup is depicted in \fg{fig1}. As in most all-optical studies of spin-coherence, we polarize electron spins using circularly polarized pump pulses (pulse length $3~\unit{ps}$ with a typical spectral width of 4~meV at 3.36~eV laser energy) and monitor their orientations during spin precession with linearly polarized probe pulses. Whereas both laser pulses are usually launched from a single laser source, we use two independently tunable Ti:Sapphire lasers, which are synchronized by a lock-to-clock system with temporal jitter less than 1~$\unit{ps}$. This setup allows for an electronically controlled delay between pump and probe pulses thus extending the maximum pump-probe-delay to the laser repetition period. With the pulse repetition frequency of the Ti:Sapphire lasers of $80\unit{MHz}$ electronic delays up to $12.5\unit{ns}$ can be achieved.

Both pump and probe pulses are first passed through a second harmonic generator (SHG) to reach the near UV energy range which is needed for optical absorption in the large band gap semiconductor ZnO with $E_{\rm G}=3.4\unit{eV}$. Both pulse trains are linearly polarized and can independently be attenuated to a variable power. Circular polarization of the pump pulses are obtained by a $\lambda/4$ wave plate. For all time-resolved measurements, typical time-averaged laser powers are $6\unit{mW}$ and $1\unit{mW}$ for pump and probe pulses, respectively.

The sample is placed in a liquid helium bath cryostat and an external magnetic field of $250\unit{mT}$ is applied perpendicular to the laser axis (Voigt geometry). The pump pulse is blocked behind the cryostat, while the probe pulse is analyzed for its Faraday ellipticity $\eta_{\rm F}$ using a $\lambda/4$ wave plate, a polarizing beam splitter and a balanced photodetector. The detector is balanced at zero spin polarization (pump blocked) and the pump-induced ellipticity $\eta_{\rm F}  = \text{arctan} (I_{\sigma^+}/I_{\sigma^-})- \pi/4$ is measured with $I_{\sigma^+}$ and $I_{\sigma^-}$ being the intensities of the right and left circularly polarized waves, respectively. Both, pump and probe pulses are modulated by optical choppers at $400\unit{Hz}$ and $6\unit{kHz}$, respectively. The voltage signal from the photodetector is processed by two lock-in amplifiers, demodulating first at the fast chopper frequency and then at the slow one.

To explore the effect of additional optical pumping on a previously generated coherent spin ensemble, we can extend our setup by a second pump pulse which is split from the probe pulse train (see dashed line in \fg{fig1}). This second pump pulse is delayed by a 3 ns mechanical delay line and then polarized and attenuated similar to the first pump pulse. In these pump-pump-probe experiments, the different beams are separated horizontally by $5\unit{mm}$ before being focused onto the sample. Here, the probe beam is in the middle and hits the sample under normal incidence. A modulation of the second pump pulse and consequently a demodulation of the Faraday ellipticity using three lock-in amplifiers in series is optional and discussed where used.

Although we believe that most of the effects presented here can be found in many semiconductors, we focus on two ZnO thin film samples which were grown by laser molecular beam epitaxy (laser MBE)\cite{Opel_arxiv} and by plasma-assisted molecular beam epitaxy (PAMBE). Sample A is a 130~nm thick epitaxial ZnO film grown on a c-plane sapphire substrate by laser MBE. Sample B was grown by PAMBE and has a thickness of 1000 nm. Prior to the ZnO deposition, a 10nm thick MgO buffer layer was deposited for the latter. X-ray diffraction shows that both samples are of high crystalline quality, as evident from the full width at half maximum (FWHM) of the ZnO(0001) reflections, which are $0.03^\circ$ for sample A and $<0.06^\circ$ for sample B. From photoluminescence (PL) measurements it is evident that sample A contains aluminum dopants while sample B is doped with aluminum and indium donors.

As demonstrated in \cite{eigenes,eigenes2} the donor electrons of the aluminum and indium dopants in these sample can selectively be polarized if the photon energy of the laser pump pulses are tuned into resonance with the respective donor-bound excitons. The angular momentum of the spin polarized electron of the exciton is then transferred to the donor electrons which leaves behind spin coherent donor electrons after exciton recombination. Spin precession of these coherent donor spin states can be detected if the probe energy is tuned close to the bound exciton energy of the same donor species. While the spin dephasing times of the respective donor spins are up to 15~ns at 10~K, we found additional mobile spins which are simultaneously excited exhibiting $T_2^*$ values of a few ns at 10~K. Although both spin states can be identified either by RSA or by conventional delay line scans, respectively, we will demonstrate in the following that their superposition can easily yield erroneous values for $T_2^*$ in time-domain delay line measurements.

In contrast to most experiments, we use the Faraday ellipticity $\eta_F$ for spin detection as it provides the largest amplitude at the exciton transition energies. While the more commonly used Faraday rotation $\theta_F$ works equally well, it exhibits a far more complicated energy dependence \cite{PhysRevB.82.155325}. For most of the experiments, we will tune both lasers to the aluminum-bound excitons from the A valence band ($D_{\text{Al}}^0X_A$, $E = 3.3608\unit{eV}$ \cite{PhysRevB.82.115207}, $\lambda = 368.9\unit{nm}$). One exception will be used for data presented in \fg{fig4}, where we achieve sign reversal of $\eta_F$ for the donor spins using optical selection rules. This is depicted in \fg{fig2} for optical excitation with right circularly polarized light from the A (\fg{fig2}a) and B valence band (\fg{fig2}b),  ($D_{\text{Al}}^0X_B$, $E = 3.3652\unit{eV}$ \cite{PhysRevB.82.115207}, $\lambda = 368.4\unit{nm}$) into the respective exciton states. For temperature dependent measurements we change the laser excitation energy by the temperature-dependent Varshni-shift of the band gap \cite{Varshni1967149} as analyzed from temperature dependent PL data (not shown).

\begin{figure}[t]%
\includegraphics*[width=\linewidth]{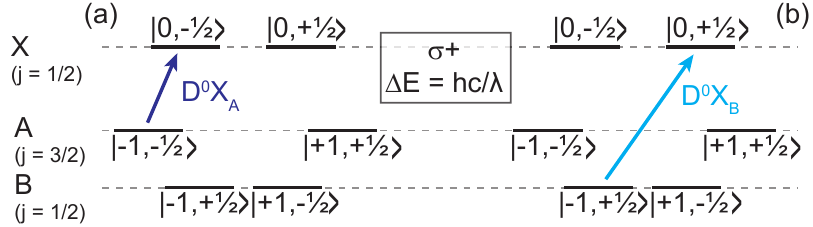}
\caption{Optical selection rules for (donor-bound) excitons in ZnO. The spin polarization changes sign when the laser energy is tuned from A to B valence band.}
\label{fig2}
\end{figure}

\section{Spin dephasing times in time-resolved Faraday ellipticity measurements}
 Typical time-resolved magneto-optical setups consist of a single pulsed laser source. Its pulse train is split into pump and probe pulses. One pulse train is temporally delayed with respect to the second one by changing its optical path length using a mechanical delay line with an optical retro-reflector on a linear stage. This setup has two disadvantages: (I) Laser pump and probe energies cannot be changed individually since they originate from the same laser source and (II) the maximum pump-probe delay is limited by the length of the mechanical delay line. There are rather long delay lines available and a single beam can in principle be run multiple times through the same delay line to achieve even longer delays, but this introduces experimental difficulties as the beam has to be adjusted more precisely to avoid delay-dependent spatial drifts. Therefore, in most optical pump-probe experiments, the maximum delay is limited to about $3\unit{ns}$, which is sufficient for most semiconductor experiments.

\subsection{Analysis of $T_2^*$ for 3~ns delay line scans}

\begin{figure}[tb]%
\includegraphics*[width=\linewidth]{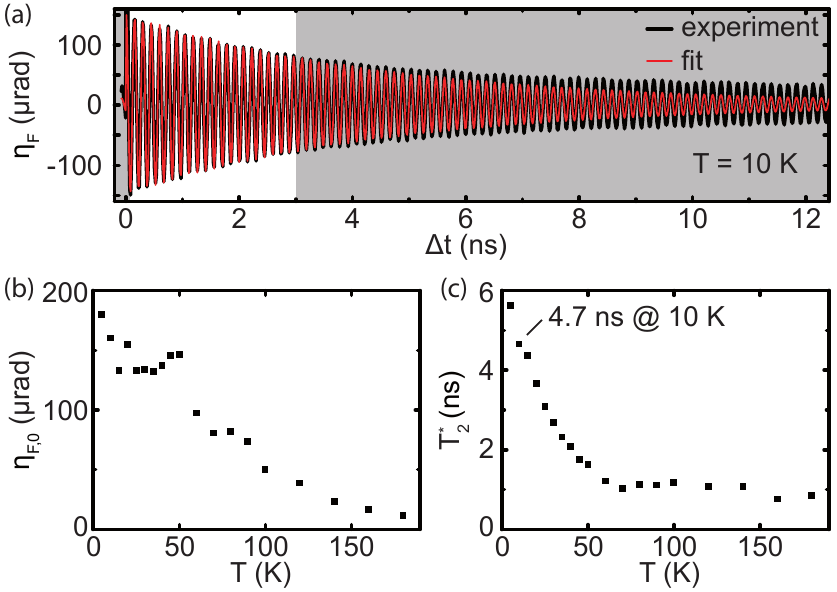}
\caption{(a) Time-resolved ellipticity (black curve) for ZnO layer (sample A) taken at $T=10$~K and $B=0.25$~T. Red curve is fit for first $3\unit{ns}$. For $\Delta t>3$~ns, it decays faster than the experimental data. (b) Amplitude $\eta_{F,0}$ from fit vs $T$, (c) $T_2^*$ from fit vs $T$.}
\label{fig3}
\end{figure}

The maximum electronic delay $\Delta_t$ in our experiment is limited by the laser repetition rate to a period of $12.5\unit{ns}$. In \fg{fig3}(a) a typical delay scan for sample A is shown after excitation of donor-bound excitons from the A valence band ($D_{\text{Al}}^0X_A$, $E = 3.3608\unit{eV}$) at $T=10$~K and $B=250$~mT. Larmor precessions extend over the full delay time with a decaying amplitude during precession. This decay is first analyzed for $\Delta_t<3$~ns, which sets the available time scale in mechanical delay line scans (see white background in \fg{fig3}(a). We fit these data by an exponentially damped cosine function
\begin{equation}
\label{delayFit}
\eta_F(\Delta t) = \eta_{F,0}\cdot\cos(\omega_L\Delta t + \delta)\exp(-\frac{\Delta t}{T_2^*})+y_0
\end{equation}
with amplitude $\eta_{F,0}$, Larmor frequency $\omega_L$ and pump-probe delay $\Delta t$. Additionally, we allow for a phase $\delta$ and an offset $y_0$. This way we can determine $T_2^*$. As seen in \fg{fig3}(a) (red curve), this fit works quite very well for $0<\Delta t<3$~ns. The deviations at longer delays ($\Delta >3$~ns) will be discussed in the next section. We measured TR ellipticity up to 180~K and used the same fitting procedure for all temperatures. In \fgs{fig3}(b) and (c) we plot the temperature dependent  $\eta_{F,0}$ and $T_2^*$, respectively. Both decrease with increasing temperature. The spin dephasing times  in \fgs{fig3} show a strong decrease at lower temperatures as seen in several semiconductor systems \cite{PhysRevLett.80.4313,PRB63_Beschoten2001_SpinCoherenceandDephasinginGaN} indicating a strong thermally activated spin dephasing mechanism which fades away at 60~K.

\subsection{Analysis of $T_2^*$ for 12.5~ns delay line scans}

\begin{figure}[bt]%
\includegraphics*[width=\linewidth]{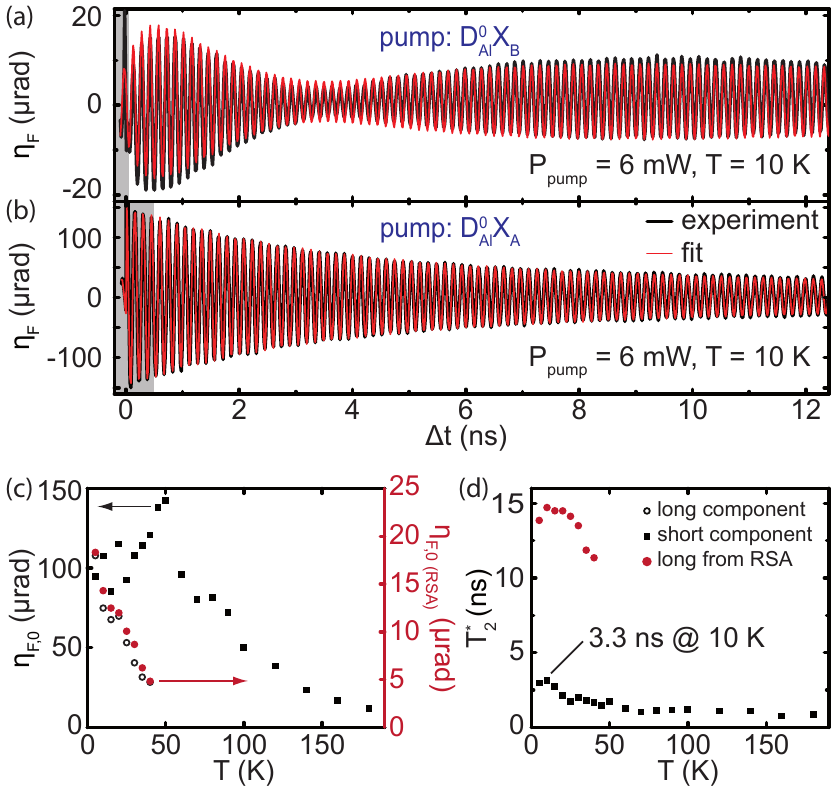}
\caption{Time-resolved ellipticity for an epitaxial ZnO film (sample A) taken at $T=10$~K and $B=0.25$~T for (a) two-color measurement (black curve with pump pulse at $D_{\text{Al}}^0X_B$ and probe pulse at $D_{\text{Al}}^0X_A$). The red curve is a fit to the 3 spin components (see text for further detail). (b) Fit of \eq{doubleFit} to $\eta_F(\Delta t)$ over the full laser repetition period (red curve) for single color experiment (black curve) taken at $D_{\text{Al}}^0X_A$. (c) Amplitudes from ellipticity fits vs $T$ for the long spin component III (open circles) and the short spin component II (squares). For comparison the amplitudes from RSA (filled red circles) are added using a different scale (axis to the right). (d) $T_2^*$ vs $T$ for short component II (filled squares) and long component III (filled red circles). The latter are extracted from RSA data (cp. to \fg{fig5}(d)).}
\label{fig4}
\end{figure}

Surprisingly, the temperature dependence of $\eta_{F,0}$ and $T_2^*$ changes completely if the full data set over the repetition period of 12.5~ns is considered. In \fg{fig3}(a) it becomes obvious that the above fits do not match the data for $\Delta t > 4 \unit{ns}$ as the sample sustains spin precession much longer than expected. We note that the single exponential fit fails to fit the data (black curve in \fg{fig3}(a)) over the full repetition period (not shown). To better visualize the different spin components contributing to $\eta_{F}$, we show a 2-color measurement in \fg{fig4}(a). While the probe beam is still in resonance with the $D_{\text{Al}}^0X_A$ exciton line, we changed the pump energy to slightly larger energies to allow for exciton excitation from the B valence band ($D_{\text{Al}}^0X_B$). According to the optical selection rules in ZnO, we expect a sign reversal of $\eta_{F}$ as discussed above. However, we observe a more complex spin precession pattern (\fg{fig4}(a)), which clearly cannot be represented by the single spin component in \eq{delayFit}. It consists of at least three oscillating spin components (red curve is fit to the three oscillating signals): (I) a very short one which decays within the exciton lifetime of a few $100\unit{ps}$) \cite{eigenes2}, (II) a short component with spin dephasing times of a few ns which stems from mobile carriers \cite{eigenes2}, and (III) a long one which decays over 15~ns, which was attributed to $T_2^*$ from the donor spins \cite{eigenes}. All three components become visible in the beating pattern as only components (I) and (III) show the expected sign reversal. The complicated polarization mechanisms of all components is not in the focus of this paper and has been discussed elsewhere \cite{eigenes,eigenes2}. We note, however, that there is only a very weak dependence of the respective spin dephasing times of all three components when the laser excitation energy is changed from the A to the B valence band \cite{eigenes2}.

In the following, we will focus on the short (II) and long component (III). As component (I) is only seen during the first few spin precession periods, we simplify our data analysis by fitting from $400\unit{ps}$. We thus fit the data in \fg{fig3}(a) with two exponentially decaying spin components by

\begin{align}
\label{doubleFit}
\eta_F(\Delta t) &= \eta_{F}^{\text{short}}\cdot\cos(\omega_L^{\text{short}}\Delta t + \delta_{\text{short}})\exp(-\frac{\Delta t}{T_{2,\text{short}}^*})\nonumber\\
&+\eta_{F}^{\text{long}}\cdot\cos(\omega_L^{\text{long}}\Delta t + \delta_{\text{long}})\exp(-\frac{\Delta t}{T_{2,\text{long}}^*})+y_0,
\end{align}

which assumes two independently decaying spin species. The fit (see \fgs{fig4}(b)) now perfectly matches the experimental data over the full laser repetition period. Within the precision of our experiment both components precess at the same frequency $\omega_L^{\text{short}} = \omega_L^{\text{long}}$ with Land\'{e} factor of $g = 1.97$. We note that $T_{2,\text{long}}^*$ for the long component cannot be fitted reliably as its value exceeds $T_{\text{rep}}$. Therefore, this value is taken from resonant spin amplification (RSA) measurements (see section 4) as an input parameter for delay line fits. The temperature dependent amplitudes and $T_2^*$ values are given in \fgs{fig4}(c) and (d), respectively. Red filled circles correspond to values taken from RSA.

Interestingly, these results tell a completely different story than those given in \fgs{fig3}(b) and (c). The long living spin component (III) with $T_2^* = 14.7\unit{ns}$ (at $T = 10\unit{K})$ only shows a rather weak temperature dependence. Its amplitude decreases almost linearly with temperature (see red circles in \fg{fig4}(c) for RSA amplitudes and black open circles for amplitudes from delay line scans) and vanishes at about $T = 50 \unit{K}$. Above $T = 50\unit{K}$ only the short component II remains. In this temperature range ($T>50$~K), both its amplitude and $T_2^*$ do not differ from previous results in \fgs{fig3}(b) and (c). Below $T = 50\unit{K}$, however, its amplitude increases smoothly with decreasing temperature and the slope of $T_2^*(T)$ does no longer strongly change and seems to follow a single spin dephasing mechanism for low and high temperatures. At first, the increasing amplitude between 5 and 50~K of the short component (see black squares in \fg{fig4}(c)) may not appear reasonable, but the data exhibits a simple process: The electron spins showing a long $T_2^*$ can be thermally activated into states of short $T_2^*$, decreasing the amplitude of the former and thus increasing the amplitude of the latter. We conclude that the existence of spin component III with long $T_2^*(T)$ was hidden in the time-resolved data over the shorter delay of 3~ns (see \fg{fig3}(a)). However, as this component has a comparable amplitude as the shorter component II at low temperatures (see \fg{fig4}(c)), it artificially enlarges $T_2^*(T)$ for the latter component when fitting over the 3~ns delay only (cp. \fg{fig4}(c) with black squares in \fg{fig4}(d)).

\section{Spin dephasing times from resonant spin amplification}
As seen above, it is difficult to extract $T_2^*$ from time-domain experiments, if one spin component exceeds the available pump-probe delay, in particular if an additional short spin component dominates the first few nanoseconds. As in our case the available pump-probe-delay equals the laser repetition period, the long component also fulfills  $T_2^* > T_{rep}$ and we can therefore additionally use the method of resonant spin amplification (RSA) \cite{PhysRevLett.80.4313}. In this technique, $\eta_F$ is measured as a function of the external magnetic field at a fixed pump-probe delay. Whenever $T_{\text{rep}}$ is a multiple of the spin precession period, the excited spins are aligned in phase which results in distinct resonances in $\eta_{F}$. The width of these resonances is a measure for $T_2^*$. Long $T_2^*$ values are seen as very sharp resonances.

The RSA data $\eta_F(B)$ can be fitted as a sum of multiple repetitions of \eq{delayFit}, with each repetition additionally delayed by the sum of $n$ subsequent repetition periods $\Delta t + nT_{\text{rep}}$ \cite{PhysRevLett.80.4313}. In order to keep the presented equations readable, we will not show $\delta$ and $y_0$ although both parameters have been used in the actual fit. For the same reason, we will only discuss the long living spin component (III) setting $T_2^* \equiv T_{2,\text{long}}^*$ and likewise for all other variables. This approach is justified by our choice of $\Delta t = 12.48\unit{ns}$ (where the probe pulse hits the sample $20\unit{ps}$ before the following pump pulse), which is about five times the spin dephasing time of the short component (II). RSA is measured as a function of $B$, which results in a linear increase of the Larmor frequency $\omega_L(B)$. In the following, we thus describe the data as a function of $\omega_L$ at fixed $\Delta t$:

\begin{align}
  \eta_F(\omega_L) &= \sum\limits_{n = 0}^{\infty}\eta_{F,0}\cdot\cos\left(\omega_L(\Delta t+nT_{\text{rep}})\right)e^{-\frac{\Delta t+nT_{\text{rep}}}{T_2^*}}\\
                   = \eta_{F,0}&e^{-\frac{\Delta t}{T_2^*}}\frac{\cos(\omega_L(\Delta t-T_{\text{rep}}))-\cos(\omega_L\Delta t)e^{\frac{T_{\text{rep}}}{T_2^*}}}{2\cos(\omega_LT_{\text{rep}})-e^{\frac{T_{\text{rep}}}{T_2^*}}-e^{-\frac{T_{\text{rep}}}{T_2^*}}}
      \label{RSAsimpel}
\end{align}

\begin{figure}[tb]%
\includegraphics*[width=\linewidth]{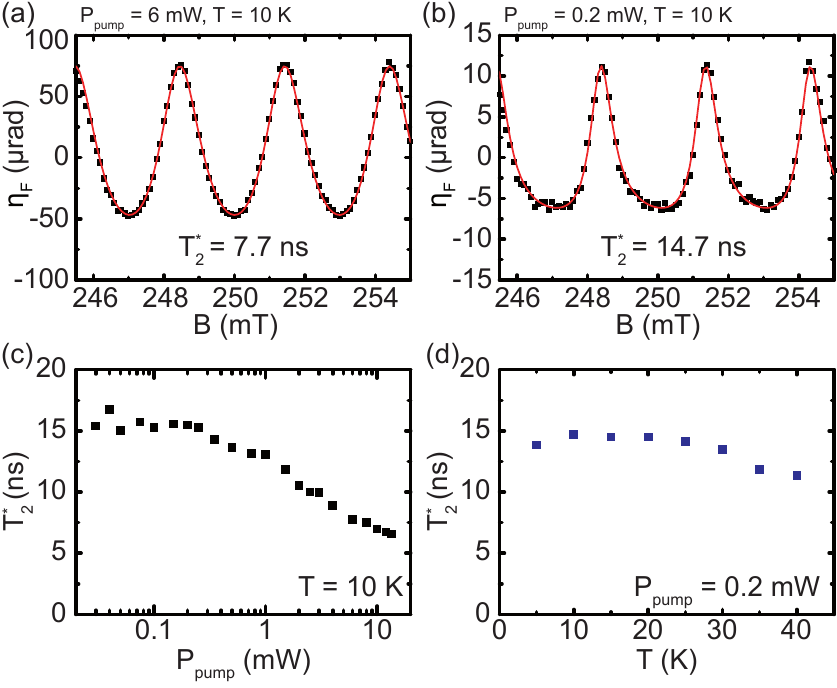}
\caption{Ellipticity from RSA scans on an epitaxial ZnO film (sample A) taken at $T=10$~K for (a) large pump power $P_{\text{pump}} = 6\unit{mW}$ and (b) low pump power $P_{\text{pump}} = 200\unit{\mu W}$. The red curves are fits to the RSA formula in \eq{RSAsimpel} (c) $T_2^*$ from RSA vs laser pump power $P_{\text{pump}}$. Long $T_2^*$ values are only observed for $P_{\text{pump}}<0.4$~mW (d) $T_2^*$ from RSA vs $T$ at low $P_{\text{pump}} = 200\unit{\mu W}$}
\label{fig5}
\end{figure}

\subsection{Inconsistent spin dephasing times from resonant spin amplification measurements}

In \fg{fig5}(a), we show an RSA measurement (black squares) on sample A at $T=10$~K. This data has been taken with the same laser parameters as for \fg{fig3}(b) ($P_{\text{pump}} = 6\unit{mW}$). In contrast to the above time-resolved scans it is measured as a function of $B$ (in the range of the previously used $B = 250\unit{mT}$) at a fixed pump probe delay $\Delta t = -20\unit{ps}$, which equals $\Delta t = 12.48\unit{ns}$ from the previous laser pump pulse. The RSA resonances are clearly resolved. Their distance corresponds to one Larmor precession cycle. We also include a fit to Eq.~4 (see red curve in \fg{fig5}(a)).

The RSA model fits the data very well, but surprisingly, the extracted spin dephasing time of $T_2^* = 7.7\unit{ns}$ is inconsistent with the above value obtained in time-domain experiments. This value neither matches the results of the single-component fit ($4.7$~ns in \fg{fig3}(c)) nor is it long enough to account for the slowly decaying spin component III in the two-component fitting. To add to the inconsistency, $T_2^*$ shows a strong dependence on the laser pump power $P_{pump}$ as shown in \fg{fig5}(c). While $T_2^*$ is around 15~ns for $P_{pump}<0.4$~mW (see RSA scan in \fg{fig4}(b) at $P_{pump}=0.2$~mW (black squares) and respective fit (red curve)), it decreases significantly at larger pump power and reaches 6.6~ns at 13.5~mW. A similar power dependence has previously been observed in GaAs and GaN at low temperatures and has been attributed to inhomogeneous spin dephasing \cite{PhysRevLett.80.4313, PRB63_Beschoten2001_SpinCoherenceandDephasinginGaN}. However, it is important to emphasize that we do not observe such a strong power dependence in the time-domain data (not shown). In delay-scans we find that at powers above $1\unit{mW}$, the ratio $\eta_F^{\text{II}}/\eta_F^{\text{III}}$ changes in favor of the short component (not shown), which we in fact attribute to laser heating as this is consistent with the temperature dependence in \fg{fig4}(c). But the  decrease in $T_2^*$ starting above $300\unit{\mu W}$ is clearly not evident in delay scans. In the following, we will show that the decrease of $T_2^*$ is not due to spin dephasing but rather results from the influence of consecutive pump pulses which can significantly diminish the remaining spin polarization from the previous pump pulses. When using the standard RSA formula in Eq.~4, this reduction of spin polarization is not taken into account but is rather misinterpreted as a decrease in $T_2^*$.

\subsection{Influence of subsequent pump pulses on spin precession}

A basic assumption for the interpretation of RSA data is the superposition of independent spin ensembles from the pump pulses of each repetition period. By extending our measurements to a pump-pump-probe setup, we will demonstrate that this assumption does not hold. In this experiment we use an additional  second pump pulse (``pump~2''), which generates a second spin ensemble. This spin ensemble can independently be time-delayed relative to the first spin ensemble generated by pump pulse~1.

In our setup, this second pump pulse is split from the probe pulse train from laser 2 and timed by a mechanical delay line (see dashed optical path in \fg{fig1}). Both lasers and hence all three pulses are tuned to the same photon energy (resonance at $D_{\text{Al}}^0X_A$). Both pump pulses are circularly polarized. While pump~2 is not modulated, pump~1 and the probe are modulated as before. Again, the signal undergoes a two-step demodulation using two lock-in amplifiers at the respective modulation frequencies of pump~1 and the probe, which effectively filters the contribution of the unmodulated pump~2. Of course, spins are still polarized by pump~2. After demodulation the measured $\eta_F$ is thus given as the difference between a delay-line scan with signals from both pump pulses and a delay scan with pump~2 only. For the latter pump~1 is blocked by its chopper. As a result, we probe the precessing spin polarization triggered by pump~1 under the influence of pump~2.

\begin{figure}[tb]%
\includegraphics*[width=\linewidth]{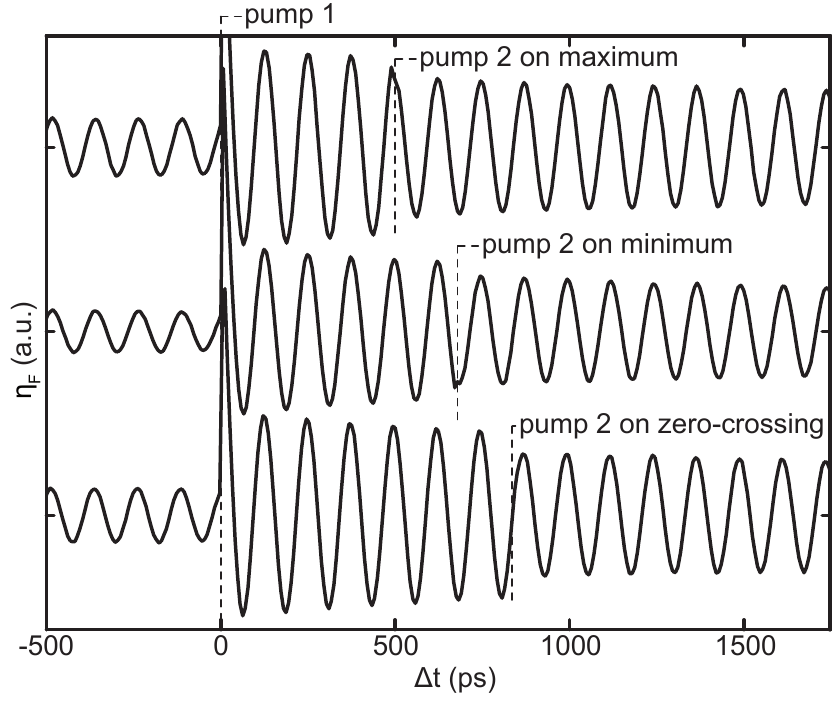}
\caption{Time-resolved ellipticity in pump-pump-probe experiment on an epitaxial ZnO film (sample A)
taken at $T = 10$\,K and $B = 0.25$\,T. The first pump pulse polarizes electron spins at $\Delta t = 0$ while a second pump pulse adds a new spin packet at different delays (see dashed lines). The latter spin packets cannot be observed directly as the second pump pulse is not optically modulated. Its influence on the spin packet which was generated by pump pulse 1 is apparent as the precession amplitude is reduced immediately.}
\label{fig6}
\end{figure}

The influence of pump~2 on the spin polarization generated by pump~1 is summarized in \fg{fig6}. At $\Delta t = 0$, the first spin ensemble is polarized by pump~1. The measured $\eta_F$ is positive and spins start to precess with a cosine function indicating that the spins are initially oriented parallel to the laser beam direction. Pump~2 hits the sample after some precession periods of the first spin packet. As pump~2 has the same photon energy and light helicity, it will create spin packets with the same initial spin orientation as for the first spin packet independent of the pump~1/pump~2 delay. By changing this delay, however, we can control the relative spin orientations of both spin packets. In \fg{fig6} we depict a series of measurements (from top to bottom) for parallel, antiparallel and perpendicular alignments of these spin packets (see dashed line). The only influence of pump~2 is an overall decrease of $\eta_F$ for all measurements. We note that there is neither a change in the precession frequency nor is the phase influenced by pump~2. If the generation of both spin packets were completely independent, however, we would not expect to observe any changes from pump~2.

As we reported in Ref. \cite{eigenes}, spin-polarized electron-hole pairs are firstly excited after optical absorption by pump~1. These exciton states are short-lived. The angular momentum of their electron spins is transferred to donor electrons of either aluminum or indium dopants. Optical absorption of pump~2 polarizes a new set of spin polarized excitons. Part of the previously polarized donor spins might be replaced during the subsequent spin transfer process. This partial loss of spin polarization from pump~1 is seen as an amplitude drop of $\eta_F$ (see \fg{fig6}) whenever pump~2 hits the sample.

\begin{figure}[tb]%
\includegraphics*[width=\linewidth]{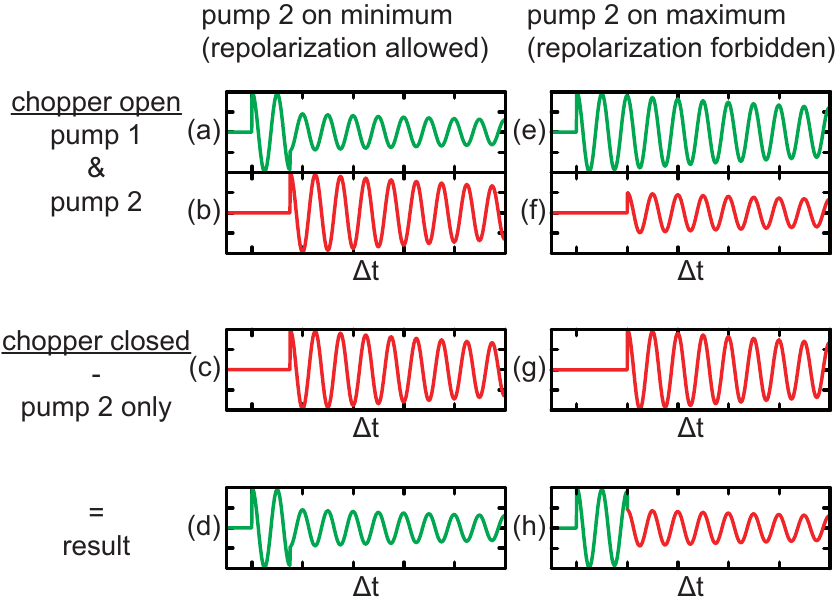}
\caption{Illustration of the ellipticity versus delay time dependence expected for different moments of pump 2. The situation is shown for the ZnO-specific case of phase dependent spin replacement.}
\label{fig7}
\end{figure}

On a closer look at this polarization model \cite{eigenes}, the efficiency of the spin replacement depends strongly on the phase (i.e. the spin orientation) of the first spin packet created by pump~1. The spin polarization can only be transferred to the donor electron if the spin polarized electron of the exciton can form a spin singlet state (i.e. antiparallel alignment) with the donor electron. Therefore, the idea of spin replacement only makes sense if pump~2 occurs at a minimum of $\eta_F$. This is illustrated in \fg{fig7}, where the respective spin states from both pump pulses are schematically depicted for different states of optical modulation. If the mechanical chopper of pump pulse~1 is open, the spins generated by pump~1 precess until some of them are repolarized by pump~2 at the minimum of $\eta_F$ (see green curve in \fg{fig7}(a)). At the same time, spin precession from the pump~2 spin packet can also be detected (see red curve in  \fg{fig7}(b)). If, however, the pump~1 chopper is closed, only the spin polarization of pump~2 is detected (\fg{fig7}(c)) with the same amplitude as in \fg{fig7}(b). After lock-in amplification, both pump~2 signals (chooper open - chopper closed) cancel out and only the pump~1 signal with the spin replacement is visible (\fg{fig7}(d)).

In contrast, no replacement is allowed if pump~2 occurs at a maximum of the pump~1 polarization. Spin precession of the pump~1 spin packet is thus not influenced by pump~2 for that case (see \fgs{fig7}(e) and (f)). But as the polarization of these donor electrons cannot be replaced, the amplitude of the polarization from pump~2 is reduced (\fg{fig7}(f)). If the chopper is closed, however, pump~2 meets unpolarized donor electrons and optical excitation is more efficient (amplitude in \fg{fig7}(g) is larger than in \fg{fig7}(f)). For lock-in detection (chopper open - chopper closed), the two pump~2 signals do not cancel out anymore and the larger signal in \fg{fig7}(g) dominates. During lock-in amplification, the latter will be subtracted from the total signal with the chopper being open (sum of \fgs{fig7}(e) and (f)). As both spin packets (green and red) are aligned parallel, this again leads to a reduced spin precession amplitude at the incidence of pump~2 (\fg{fig7}(h)).

We believe that the observed spin replacement is not specific to ZnO and might also be seen in other systems. If, for example, electron spins are optically pumped in n-GaAs (or any n-doped direct band gap semiconductor), electrons are excited from the valence into the conduction band states. Additionally, holes are created in the valence band, which will eventually recombine with the conduction band electrons. Since we are considering n-doped samples, there will be plenty of electrons in the conduction band, which have not been polarized by the laser pump pulse, but still may recombine with the holes. Similarly, if these electrons have been polarized by a previous pump pulse, they may recombine with holes generated by a following pump pulse. During that process, the polarization of a pump pulse is reduced (i.e. replaced) when the next pump pulse polarizes a new subset of electron spins.

The above spin replacement scheme explains the observed decrease of $\eta_F$ at the arrival of pump 2 no matter if pump~2 coincides with a maximum, minimum or zero-crossing of the precession initiated by pump~1, as indeed seen in \fg{fig6}.

\subsection{Correct interpretation of spin dephasing times from RSA}

\begin{figure}[tb]%
\includegraphics*[width=\linewidth]{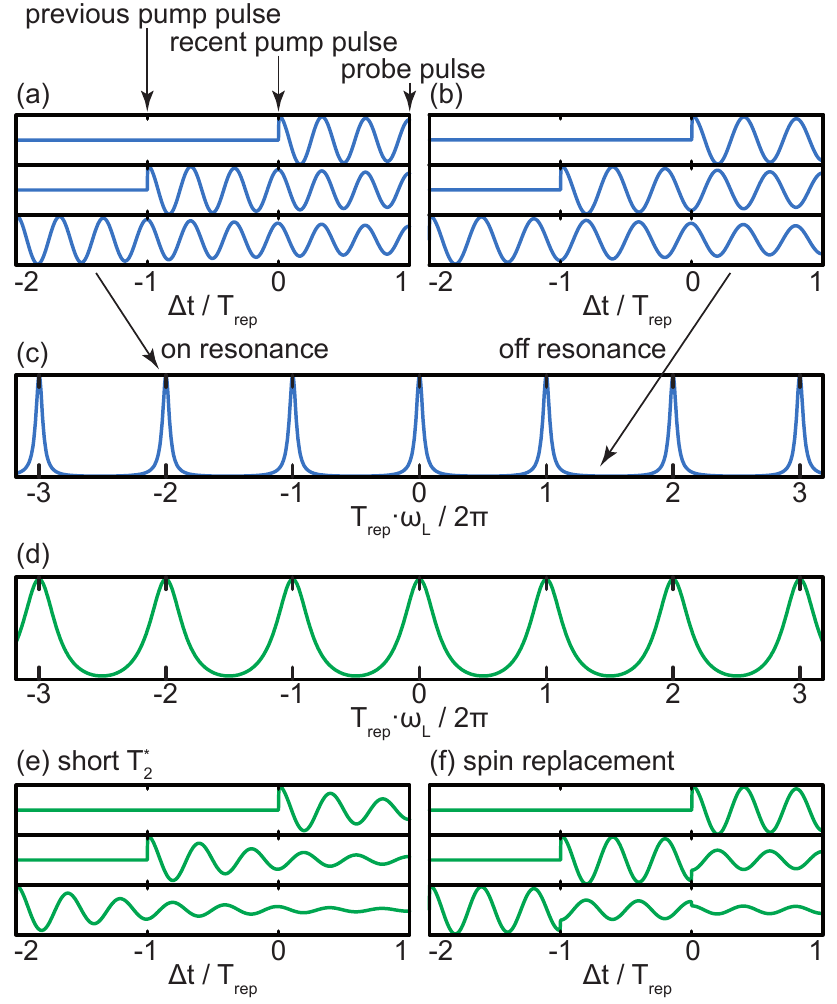}
\caption{Illustration of the RSA effect for (a) long $T_2^*$ on resonance and (b) long $T_2^*$ off resonance, leading to an RSA signal with narrow peaks (c). These peaks become broader when (d) $T_2^*$ is reduced as shown in (e) for an off-resonance example. (f) The replacement of spins by subsequent pump pulses has the same effect.}
\label{fig10}
\end{figure}

We now discuss the effect of spin replacement on the RSA measurements. To better understand its influence on the spin dephasing times obtained from RSA, we first explore how $T_2^*$ changes the RSA resonances if no spin replacement is present. In \fg{fig10}(a) spin precession of three spin ensembles from three subsequent laser pump pulses are shown. In this first example, $T_{\text{rep}}$ is a multiple of the spin precession period $2\pi/\omega_L$, so the spin amplitudes from subsequent pulses add up to an amplified signal. However, in the case of \fg{fig10}(b) spin polarization from each laser pump pulse has the opposite sign relative to the polarization from the previous pump pulse. Each newly added spin packet thus cancels with the polarization from the previous spin packet yielding a small ellipticity signal. For long $T_2^*$ values more than the illustrated three periods have to be taken into account and therefore $\omega_L$ and thereby $B$ have to be tuned very accurately to achieve maximum amplification, which is seen as narrow resonances in \fg{fig10}(c).

On the other hand, if $T_2^*$ is shorter (\fg{fig10}(e)) fewer periods have to be taken into account. The amplification is weaker, but more importantly, $B$ does not need to be tuned as precisely, as only few spin packets have to be considered leading to broader resonances (\fg{fig10}(d)). The broader resonances are also seen if the spin polarization is partially replaced by subsequent laser pulses. For a single spin packet with long $T_2^*$, its amplitude is reduced each time a new spin packet is generated. Essentially, after a few laser periods, its amplitude is reduced as if it had a short $T_2^*$ to start with (compare \fgs{fig10}(e) and (f)). If spin replacement is not considered, the extracted spin dephasing times are thus much shorter than their actual values.

Mathematically, the spin replacement can simply be modeled as an additional factor $\alpha$ for each additional pump pulse which occurs during the total lifetime of the spin ensemble. In other words, $\alpha$ determines the amount of spins which remains from the previous pump pulse:

\begin{align}
  \eta_F(\omega_L) &= \sum\limits_{n = 0}^{\infty}\alpha^n \eta_{F,0}\cdot\cos\left(\omega_L(\Delta t+nT_{\text{rep}})\right)e^{-\frac{\Delta t+nT_{\text{rep}}}{T_2^*}}\\
                   = \sum\limits_{n = 0}^{\infty}&\eta_{F,0}\cdot\cos\left(\omega_L(\Delta t+nT_{\text{rep}})\right)e^{-\frac{\Delta t}{T_2^*}}\left(\alpha e^{-\frac{T_{\text{rep}}}{T_2^*}}\right)^n\\
                   = \eta_{F,0}&e^{-\frac{\Delta t}{T_2^*}}\frac{\alpha\cos(\omega_L(\Delta t-T_{\text{rep}}))-\cos(\omega_L\Delta t)e^{\frac{T_{\text{rep}}}{T_2^*}}}{\alpha2\cos(\omega_LT_{\text{rep}})-e^{\frac{T_{\text{rep}}}{T_2^*}}-\alpha^2 e^{-\frac{T_{\text{rep}}}{T_2^*}}}
      \label{RSAalpha}
\end{align}

It can easily be seen, that $\alpha$ can be expressed by an apparent spin dephasing time $T_{\text{app}}$ with

\begin{align}
& e^{-\frac{T_{\text{rep}}}{T_{\text{app}}}} := \alpha e^{-\frac{T_{\text{rep}}}{T_2^*}}\label{alphaEqn}\\
\Rightarrow & T_{\text{app}} = \frac{T_{\text{rep}}}{\frac{T_{\text{rep}}}{T_2^*}-\ln\alpha}.
\end{align}

If we apply this relation to \eq{RSAalpha}, we get exactly the same fit function as \eq{RSAsimpel} with only a different amplitude.

\begin{figure}[tb]%
\includegraphics*[width=\linewidth]{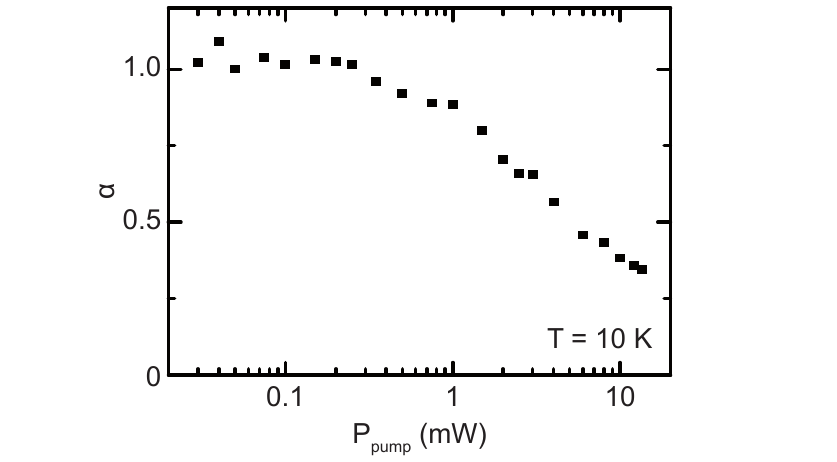}
\caption{Factor $\alpha$ calculated from \fg{fig5}(c) using \eq{alphaEqn} under the assumption of a constant $T_2^* = 15\unit{ns}$. Note, that this assumption probably neglects some additional dephasing, so that $\alpha$ includes more effects than just the spin replacement.}
\label{fig11}
\end{figure}

All spin dephasing times in \fg{fig5}(b) have been determined by the standard RSA analysis ignoring the  effect of spin replacement. The apparent $T_2^*$ from RSA is decreased when spins are being replaced, which becomes dominant at large photon densities. In contrast at low laser power, it becomes statistically irrelevant. As only few spins are excited by each pump in that case, it is unlikely that they have already been polarized by the previous pump pulse. We note that spin replacement is not directly visible in time-domain measurements as the remaining polarization at $\Delta t < 0$ from the previous pump pulse only influences the amplitude and phase of the subsequent polarization at $\Delta t > 0$. $T_2^*$ is, on the other hand, determined from the decay of $\eta_F$ for $\Delta t > 0$ and is therefore not subject to the influence of additional pump pulses.

From \eq{alphaEqn}, we can calculate $\alpha$ from the apparent RSA spin dephasing times $T_{\text{app}}$. In \fg{fig11} this has been done treating $T_2^*$ from \fg{fig5}(c) as $T_{\text{app}}$ assuming that $T_2^* = 15\unit{ns}=const.$ as retrieved from standard RSA analysis at low pump powers (cp. to \fg{fig5}(b)). For $P_{\text{pump}} = 6\unit{mW}$ (as shown in \fg{fig5}(a)) we obtain $\alpha = 0.46$. This means that 54~\% of the spins are being replaced during consequtive pumping at 6~mW, resulting in the apparently shorter spin dephasing time ($T_{\text{app}}^*=7.7$~ns) in \fg{fig5}(a). Spin replacement explains the inconsistent values for $T_2^*$ from RSA and from time-domain measurements and is obviously negligible for low $P_{\text{pump}}$ when $\alpha=1$.

We note, however, that in our model the probability for spin transfer from spin polarized excitons to donor electrons and hence the magnitude of $\alpha$ depends on the exact orientation of the previously polarized spin packet, which continuously changes with $\omega_L(B)$. Therefore our assumption of $\alpha$ being independent of $\omega_L$ is a simplification. Furthermore, we cannot entirely rule out additional spin dephasing from thermal effects or inhomogeneous spin dephasing. Although the observed drop in $T_{\text{app}}$ and $\alpha$ as a function of laser power is clearly dominated by spin replacement, these additional effects may also contribute, which might explain that $\alpha$ falls below 0.5 above $P_{pump}=7$~mW.

Nevertheless, we emphasize that $T_2^*$ values can be obtained from the standard RSA analysis when using  small pump powers. On the other hand, very low laser intensities give a bad signal-to-noise ratio. We therefore chose $P_{pump}=200\unit{\mu W}$ as it sets the upper bound before spin replacement becomes evident. With this setting, we are finally able to measure the temperature dependence of $T_{2,\text{long}}^*$ (\fg{fig5}(d)), which provides consistent results with time-resolved measurements.

\subsection{Spin interaction between aluminum and indium donor spin}

\begin{figure}[tb]%
\includegraphics*[width=\linewidth]{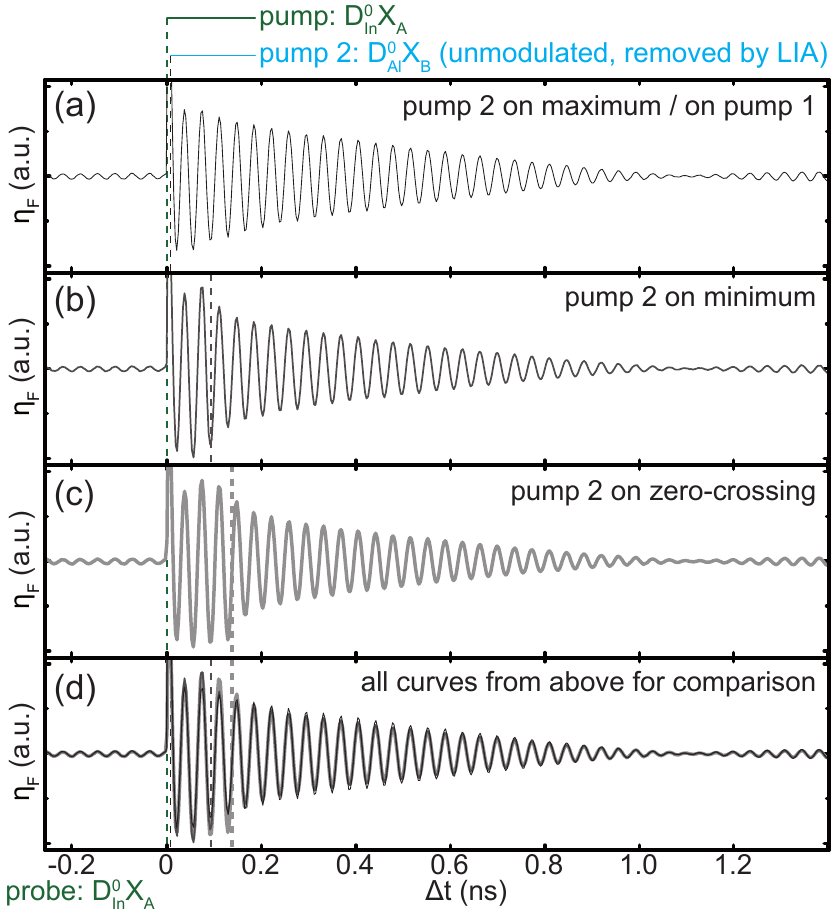}
\caption{Time-resolved pump-pump-probe experiment on a sample with different donor sites (Al and In). Pump~1 polarizes electron spins from indium donors at $\Delta t = 0$ and pump~2 is added at increasing delays from (a) to (c) as indicated by the dashed lines. (d) shows all three curves (a) to (c) for amplitude and phase comparison. Pump~2 is tuned to aluminum transitions and cannot be observed directly as it is unmodulated. The probe energy is identical to the first pump and corresponds to indium.}
\label{fig9}
\end{figure}

As described above, spin replacement in the pump-pump-probe delay scans only reduces the spin amplitude from the first spin packet, but does not change its phase or precession frequency demonstrating that there is no coherent interaction between both spin packets. We note, however, that coherent control of the magnetic exchange interaction between localized electron spins has recently been achieved in self-assembled InGaAs/GaAs quantum dots \cite{PhysRevLett.107.137402}. Due to an inhomogeneous size distribution of the self-assembled quantum dots, there is an energy broadening of about 20 meV in the energy level spectrum. It was thus possible to initialize coherent electron spins in a quantum dot subset using a first ps laser pulse. When using a second ps laser pump pulse at slightly different photon energies, two individual subsets of spins could be addressed independently. Both subsets precess in an external magnetic field. The most striking achievement was the coherent control of the relative phase of the two precessions by a time difference between the two pump pulse trains. In other words, the precession of one subset acquires a phase shift that depends on the difference in relative orientation and Zeeman energy between the two spin subsets.

As we also expect being able to trigger such a magnetic exchange interaction between localized donor spins we tested the above pump-pump-probe experiment (\fg{fig6}) on ZnO sample B. In contrast to sample A, it has aluminum and indium impurities. As demonstrated in \cite{eigenes}, we are able to selectively polarize donor spins on either aluminum or indium sites by choosing appropriate laser energies which differ by only a few meV. This way we spin polarized donor electrons on indium sites via $D_{\text{In}}^0X_A$ ($E = 3.3567\unit{eV}$ \cite{PhysRevB.82.115207}, $\lambda = 369.4\unit{nm}$) by the pump pulse 1 while spin polarized donor electrons on aluminum sites are generated by pump pulse 2 via $D_{\text{Al}}^0X_B$. We chose these exciton transitions as both subsets can be polarized exclusively since no other optical transitions are allowed at these photon energies \cite{eigenes}.

In \fg{fig9} we show a series of pump-pump-probe measurements of the Faraday ellipticity on samples B at $T=10$~K and $B=0.25$~T using the same two-step demodulation with the pump pulse 2 being unmodulated as above in \fg{fig6}. As the probe beam energy is tuned to $D_{\text{In}}^0X_A$, it only probes spin states on indium sites. As above for sample A, we do not observe any changes in either phase or precession frequency of the indium donor spins independent of the relative alignments of indium and aluminum donor spins (see top three curves in \fg{fig6} for parallel, antiparallel, and perpendicular alignment). For easier comparison all three curves are superimposed in the bottom curve of \fg{fig6}. We conclude that optical control over magnetic exchange interaction between donor spins on indium and aluminum sites is thus not feasible in our sample. We note that this is not entirely unexpected considering the low donor concentration and similar Land\'{e} factors of the shallow aluminum and indium donors \cite{PhysRevB.25.6049,Gonzalez1982357,PhysRevB.77.115334}. Interestingly though, the spin replacement even occurs across different donor species. This effect, however, seems weaker than in \fg{fig6}.

\section{Conclusions}
We have studied spin dephasing in aluminum doped ZnO by time-resolved Faraday ellipticity measurements using time-domain and $B$ field dependent resonant spin amplification scans. After a careful analysis avoiding any pitfalls and misinterpretation, reliable values for the spin dephasing times have been derived.

We demonstrated that a standard pump-probe delay scan can easily be misinterpreted, if the spin dephasing time exceeds the experimentally available pump-probe delay. In general this is obvious, but this issue can be masked by an additional spin component of shorter spin dephasing time. In this case, the measured spin precession signal may still appear to stem from a single spin component with an apparent spin dephasing time at intermediate time scales. As shown for ZnO, even a strong temperature dependence of this spin dephasing time appears to be genuine. We were able to overcome these shortcomings by using two independently tunable laser sources, which on the one hand allowed for scanning the pump-probe delay over the full laser repetition period of 12.5~ns. On the other hand, this provided independent control over pump and probe laser energies, which unambiguously proved three spin components with different spin dephasing times in time-resolved ellipticity data. While the distinction between the different spin components might be specific to ZnO, extended pump-probe delay scans can be evaluated for any material. Even if the second laser is not available, our tests for the long-lived spin component can still be carried out by adding constant delays using different optical paths or simply by having a close look to the data at $\Delta t < 0$. While some data for negative delay is routinely recorded in most experiments, it might be tempting to overlook small deviations in the fit results which may lead to completely erroneous results when analyzing spin dephasing mechanisms.

Resonant spin amplification is not sensitive to spin ensembles with a short spin dephasing time and is thus not susceptible to the same difficulties as the time-domain measurements. In ZnO, however, we also found deceptive results for spin dephasing times from RSA at larger laser pump powers. They result from the partial spin replacement by subsequent laser pulses, which seemingly reduce the spin dephasing times. This issue is again not necessarily specific to ZnO as the problem lies in the assumption, that subsequent spin polarizations are independent and add up linearly to a total polarization. We expect that measured spin replacement will probably take place in any semiconductor system, in which the spin polarization is transferred from an initial excitation (i.e. donor-bound exciton) to a different final state (i.e. donor electrons) or whenever electron spins can be reoriented by consecutive pump pulses. Its relevance can easily be tested by comparing laser power dependent RSA spin parameters to results from time-domain measurements at long delays.

\begin{acknowledgement}
This work was supported by the Deutsche Forschungsgemeinschaft (DFG)
via SPP 1285 (Projects No. BE 2441/4 and GR 1132/14).
\end{acknowledgement}

\bibliographystyle{pss}
\providecommand{\WileyBibTextsc}{}
\let\textsc\WileyBibTextsc
\providecommand{\othercit}{}
\providecommand{\jr}[1]{#1}
\providecommand{\etal}{~et~al.}

\end{document}